\documentclass[prl,twocolumn,floats,aps,epsfig,nofootinbib,amssymb]{revtex4-1}
\usepackage{graphicx}
\usepackage{cancel}
\usepackage{amssymb}
\usepackage{textcomp}
\usepackage{amsmath}
\usepackage{bm}
\usepackage{times}
\usepackage{epsfig}
\usepackage{color}
\usepackage{graphics}
\usepackage{hyperref}

\usepackage{epsfig}
\usepackage{amsmath}
\usepackage{bm}
\usepackage{times}
\usepackage{graphicx}
\usepackage{color}
\usepackage{slashed}
\usepackage{graphicx}
\usepackage{amsmath, amssymb}

\hypersetup{
    pdfnewwindow=true,      
    colorlinks=true,       
    linkcolor=black,          
    citecolor=blue,        
    filecolor=blue,      
    urlcolor=blue           
}

\def\bea{\begin{eqnarray}}
\def\eea{\end{eqnarray}}

\begin{document}

\title{\Large {\bf{Supersymmetric Dark Matter Sectors}}}
\author{Jonathan M. Arnold}
\affiliation{California Institute of Technology, Pasadena, CA 91125, USA}
\author{Pavel Fileviez P\'erez}
\affiliation{Center for Cosmology and Particle Physics (CCPP) \\
New York University, 4 Washington Place, NY 10003, USA }
\author{Bartosz Fornal}
\affiliation{California Institute of Technology, Pasadena, CA 91125, USA.}

\date{\today}
\begin{abstract}
We investigate the properties of a dark matter sector where supersymmetry is a good symmetry. 
In this context we find that the stability of the dark matter candidate is possible even when R-parity is broken 
in the visible sector. In order to illustrate the idea we investigate a simple scenario where the dark matter 
candidate is the lightest scalar field in the dark sector which annihilates mainly into two sfermions when 
these channels are available. We study the relic density constraints and the predictions for the dark matter 
detection experiments.   
\end{abstract}
\maketitle

\section{I. Introduction}
The possibility to describe the properties of the cold dark matter in the universe using 
a candidate in various particle physics scenarios has been studied for a long time. 
For a review of different candidates see Ref.~\cite{Drees:2012sz}. One of the most popular 
dark matter candidates is the lightest supersymmetric particle in SUSY theories. 
In this type of scenario typically one considers the lightest neutralino
~\cite{Goldberg:1983nd,Ellis:1983ew,Jungman:1995df} or the gravitino~\cite{Buchmuller:2009fm} 
as dark matter candidates. Both candidates have been investigated in great detail 
by many experts in the field. Unfortunately, in these models one has a large number 
of free parameters and it is difficult to make unique predictions which can be tested 
in current or future dark matter experiments. 

It is well-known that in order to guarantee the stability of the lightest neutralino in supersymmetric models 
the so-called R-parity symmetry is assumed. The case of the gravitino is different because its lifetime can be large 
enough even if R-parity is broken~\cite{Buchmuller:2009fm}. The possibility to understand the origin of R-parity 
conservation has been investigated by many groups. However, the simplest way to study this issue is to consider 
the B-L extensions of the Minimal Supersymmetric Standard Model (MSSM) where after symmetry breaking one 
can obtain R-parity as a symmetry at the low-scale. See Refs.~\cite{Krauss:1988zc,Martin:1992mq,Aulakh:1999cd,
Aulakh:2000sn,Babu:2008ep,Feldman:2011ms,Perez:2011zx} for the study of this problem in some 
supersymmetric scenarios and Refs.~\cite{Basso:2012gz,O'Leary:2011yq,FileviezPerez:2011kd} 
for recent phenomenological studies of these models. Unfortunately, even if in these scenarios 
we can understand dynamically the origin of R-parity conservation it is difficult to make interesting predictions 
for dark matter experiments since we can have several dark matter candidates, the neutralinos or right-handed sneutrinos, 
and as in the MSSM there are many free parameters.  

In this Letter we investigate the properties of a dark matter sector where supersymmetry is a good symmetry before 
the breaking of the gauge symmetry. In this context we do not need to impose a discrete symmetry to guarantee the 
stability of the dark matter candidate and even if R-parity is broken in the visible sector he dark matter candidate 
is stable. To study this idea of having a supersymmetric sector we consider a simple scenario where in the visible 
sector we have the minimal B-L extension of the MSSM~\cite{Barger:2008wn} and in the dark sector we have two 
chiral superfields with B-L quantum numbers. Here the link between the visible and dark sector is defined by the B-L 
gauge force which is broken in the visible sector by the vacuum expectation value (VEV) of the right-handed sneutrinos.
We find that after the B-L breaking a mass splitting is induced in the dark sector and the lightest field is the only 
possible candidate for the cold dark matter in the universe. In this model the dark matter candidate annihilates mainly 
into two sfermions when these channels are available. We investigate the different scenarios where we can achieve 
the observed dark matter relic density and the possible predictions for dark matter experiments. 
We find that the current bounds from the Xenon100 experiment set strong constraints on this type of models 
where the elastic dark matter nucleon cross section is through a neutral gauge boson. 

This article is organized as follows: In Section II we define a simple scenario with 
a supersymmetric dark matter sector. In Section III we show the possible scenarios 
where one can achieve the relic density observed by the experiments. The 
constraints coming from the direct detection experiments are investigated in Section IV, 
while we summarize the main results in Section V.
\section{II. Supersymmetric Dark Sector}
In general we can consider a simple extension of the standard model which is composed of 
a visible sector, a dark matter sector and the interactions between them. In this case the Lagrangian can be written as
\begin{equation}
{\cal{L}}={\cal L}_{visible} \ + \ {\cal L}_{dark} \ + \ {\cal L}_{link}.
\end{equation}
The visible sector here could be the Standard Model (SM) or any well-known extension of the SM. 
Since we are interested in the case where the dark sector is supersymmetric, one can have a 
model with broken SUSY in the visible sector but SUSY still is a good symmetry in the dark matter sector.  
In order to achieve this type of scenario we can assume that supersymmetry breaking is 
mediated as in ``gauge mediation", where the messenger fields only have quantum numbers of the visible sector, 
and the soft terms induced by gravity are very small. Then, the 
SM singlet fields in the dark sector do not get large soft terms from gravity mediation. 
In this way, we can say that supersymmetry is a good symmetry in the dark sector.

In order to illustrate this idea we will use as visible sector the simplest B-L extension of the MSSM~\cite{Barger:2008wn} 
where one can understand the origin of R-parity violating interactions. The dark sector will be composed of the chiral superfields
$\hat{X}$ and $\hat{\bar{X}}$ with B-L quantum numbers $\pm n_{BL}$. Then, the Lagrangian reads as
\begin{equation}
{\cal L}_{SDM}= {\cal L}_{B-L} \ + \  {\cal L}_{DM},
\end{equation}
where
\begin{eqnarray}
{\cal L}_{DM} &=& \int d^2 \theta d^2 \bar{\theta}  \  \hat{X}^\dagger  e^{g_{BL} n_{BL} \hat{V}_{BL}}  \hat{X} \nonumber \\
& + &  \int d^2 \theta d^2 \bar{\theta}  \  \hat{\bar{X}}^\dagger  e^{- g_{BL} n_{BL} \hat{V}_{BL}}  \hat{\bar{X}} \nonumber \\
& + &  \left(  \int  d^2 \theta \ \mu_X \hat{X}  \hat{\bar{X}} \ + \  \rm{h.c.} \right),
\end{eqnarray}
and the superpotential of the minimal B-L model is given by
\begin{eqnarray}
{\cal W}_{B-L}&=& Y_u \hat{Q} \hat{H}_u \hat{u}^c \ + \  Y_d \hat{Q} \hat{H}_d \hat{d}^c \ + \  Y_e \hat{L} \hat{H}_d \hat{e}^c \nonumber \\
& + &  Y_\nu \hat{L} \hat{H}_u \hat{\nu}^c \ + \  \mu \hat{H}_u \hat{H}_d. 
\end{eqnarray}
See Refs.~\cite{Barger:2008wn,FileviezPerez:2012mj} for the details of the minimal B-L extension of the MSSM. 
It is important to mention that there is no need to add extra Higgses in the visible sector in order to break the B-L gauge symmetry. 
In this case B-L is broken by the VEV of the right-handed sneutrinos as studied in Refs.~\cite{Barger:2008wn,FileviezPerez:2012mj}.
We will show that once the right-handed sneutrino gets a VEV R-parity is spontaneously broken but still the dark matter candidate is stable.
Here we assume that the fields, $X$ and $\bar{X}$, do not have interactions with the right-handed neutrinos, i.e. the couplings $(\hat{\nu}^c \hat{\nu}^c)^p \hat{\bar{X}}^n$ 
are not present. This means that $2p - n \ n_{BL} \neq 0$, where n and p are integer numbers.

One of the most interesting consequences of having ``exact" supersymmetry in the dark sector is that the scalar fields, $X$ and $\bar{X}$, 
do not get a VEV in most of the cases. Using the Lagrangian above we can compute the scalar potential for the $X$ and $\bar{X}$ fields, 
\begin{eqnarray}
V(X,\bar{X}) &=& |\mu_X|^2 \left(  |X|^2 \ + \  |\bar{X}|^2 \right) \nonumber \\
& + & \frac{g_{BL}^2}{8}  \left(  \frac{v_R^2}{2}  \ + \  n_{BL}  \left(  |X|^2 - |\bar{X}|^2  \right) \right)^2.   
\end{eqnarray}
Notice that here we have included the contribution to the B-L D-term due to the VEV of right-handed sneutrinos, $\left< \tilde{\nu}^c\right>=v_R/\sqrt{2}$, 
the field which breaks B-L in the visible sector. Then, one can see from the above equation that once B-L is broken we induce a mass splitting between 
the scalar fields in the dark matter sector. The relevant scalar potential for our discussion is given by
\begin{equation}
V(v_X, v_{\bar{X}} )= \frac{1}{2} M_X^2 v_X^2 \ + \  \frac{1}{2} M_{\bar{X}}^2 v_{\bar{X}}^2 \ + \  \frac{g_{BL}^2  n_{BL}^2}{32} \left( v_X^2 \ - \  v_{\bar{X}}^2  \right)^2, 
\end{equation}
where
\begin{equation}
M_X^2= |\mu_X|^2 \ + \  \frac{g_{BL}^2}{8} n_{BL} v_{R}^2, \  \  \  M_{\bar{X}}^2= |\mu_X|^2 \ - \  \frac{g_{BL}^2}{8} n_{BL} v_{R}^2,
\label{scalar-masses}
\end{equation}
and we find the following minimization conditions:
\begin{eqnarray}
\left( M_X^2 \ + \ \frac{g_{BL}^2}{8} n_{BL}^2 \left( v_X^2 \ - \  v_{\bar{X}}^2 \right) \right) v_{X}&=& 0, 
\label{minimization1} \\
\left( M_{\bar{X}}^2 \ - \ \frac{g_{BL}^2}{8} n_{BL}^2 \left( v_X^2 \ - \  v_{\bar{X}}^2 \right) \right) v_{\bar{X}} &=& 0.
\label{minimization2}
\end{eqnarray}
Now, we can think about different scenarios: 

\begin{itemize}

\item Case 1) We can have the trivial solutions, $v_{X}=v_{\bar{X}}=0$, and the lightest field in the dark sector is stable. 

\item Case 2)  $v_{X} \neq 0$ and $v_{\bar{X}} \neq 0$:
Using the Eq.(\ref{minimization1}), and Eq.(\ref{minimization2}) we can show that in this case there is a solution 
only when $\mu_X=0$. However, in this case the fermion partners $\tilde{X}$ and $\tilde{\bar{X}}$ are massless.
\item Case 3)  $v_{X} = 0$ and $v_{\bar{X}} \neq 0$:
In this case we can have the solution
\begin{equation}
v_{\bar{X}}^2 = - \frac{8 M_{\bar{X}}^2}{g_{BL}^2 n_{BL}^2}.
\end{equation}
if $M_{\bar{X}}^2 < 0$.
\item Case 4) $v_{X} \neq 0$ and $v_{\bar{X}} = 0$: There is no solution in this case.

\end{itemize}
Then, in general we can say that the scalar fields do not get a VEV even if they 
have a mass splitting due to the B-L D-term and when $M_{\bar{X}}^2 > 0$. 
This is an important result which guarantees the stability of the lightest field in the dark sector 
and we do not need to impose any extra discrete symmetry. Notice that in this analysis we have 
neglected the contribution from the kinetic mixing between hypercharge and B-L, which 
does not change our conclusion.

In order to understand the existence of a dark matter candidate let us study the spectrum in the dark matter sector. 
In Eq.~(\ref{scalar-masses}) we have the masses for the scalar fields, while the mass of the fermionic candidates 
is 
\begin{eqnarray}
{\cal M}_{\tilde{X}_1}&=& {\cal M}_{\tilde{X}_2}=\mu_X.
\end{eqnarray}
Therefore, the lightest field in the dark sector is the scalar field $\bar{X}$. Here we are using the convention where $n_{BL}$ is positive.
Now, are these fields stable at cosmological scales? 

The field $X$ can decay into its superpartner $\tilde{X}$ and SM fermions because R-parity is broken in the visible sector.
In the case of $\tilde{\bar{X}}$ and $\tilde{X}$ we have a similar situation, they can decay to $\bar{X}$ and SM fermions as well. 
Therefore, only the lightest field in the dark sector, $\bar{X}$, can be stable even if R-parity is broken in the visible sector.
This is an interesting result which is a consequence of having ``exact" supersymmetry in the dark matter sector before B-L is broken in the 
visible sector. Before we finish this section we would like to stress the existence of the relation between the masses of all fields in the dark 
sector:
\begin{equation}
M_{X}^2= M_{\tilde{X}}^2 \ + \  \frac{1}{2} n_{BL} M_{Z_{BL}}^2 = M_{\bar{X}}^2 \ + \  n_{BL} M_{Z_{BL}}^2,
\end{equation}
where the mass of the B-L neutral gauge boson is given by $M_{Z_{BL}}= g_{BL} v_R / 2$. Notice that the supertrace mass formula, 
$\rm{Str} \ M^2=0$, is valid here since we have the same splitting for the scalar components but with different signs. 
Here we neglect possible Planck scale suppressed operators due to gravity effects.
\section{III. Dark Matter Relic Density}
The B-L D-term defines how the dark matter candidate annihilates into two sfermions when these channels are available.
Here we will focus on the scenarios where the dark matter candidate is always heavier than a least one sfermion in the 
MSSM. In the case when the mass of $\bar{X}$ is below $M_{Z_{BL}}/2$ the main 
annihilation channels are in fact those with two sfermions:
\begin{equation}
\bar{X} \bar{X}^{\dagger} \  \to \  \tilde{f}_i \ \tilde{f}_i^\dagger, 
\end{equation}
and the annihilation cross section in the non-relativistic limit is given by
\begin{eqnarray}
\sigma \left( \bar{X} \bar{X}^{\dagger} \  \to \  \tilde{f}_i \ \tilde{f}_i^\dagger \right) v &=& \frac{1}{64 \pi} \frac{1}{M_{\bar{X}}^2}
\sqrt{1 \ - \  \frac{M_{\tilde{f}_i}^2}{M_{\bar{X}}^2}} \nonumber \\
&\times&  | \lambda|^2 \  \left| 1 + \frac{M_{Z_{BL}}^2}{4 M_{\bar{X}}^2 \ - \  M_{\tilde{\nu}^c}^2 } \right|^2
\end{eqnarray}
Here $M_{\tilde{\nu}^c}=M_{Z_{BL}}$ and $\lambda= g_{BL}^2 n_{BL}/4$ for sleptons. Now, we can compute the approximate freeze-out temperature 
$x_f = M_{\bar{X}}/T_f$. Writing the thermally averaged annihilation cross section as 
$\left< \sigma v \right> = \sigma_0 (T/M_X)^n$, then the freeze-out temperature is given by
\begin{eqnarray}
x_f &=& \ln\left[0.038(n+1)\left({g \over \sqrt{g_*}}\right) \ M_{Pl} M_{\bar{X}} \sigma_0\right]  \nonumber \\
&-& (n + {1\over 2}) \ln \left[\ln \left[0.038(n+1)\left({g \over \sqrt{g_*}}\right) \ M_{Pl} \ M_{\bar{X}} \sigma_0\right]\right], \nonumber \\
\end{eqnarray}
where $M_{Pl}$ is the Planck mass, $g$ is the number of internal degrees of freedom and $g_*$ is the effective number 
of relativistic degrees of freedom evaluated around the freeze-out temperature.
As is well-known, the present day energy density of the relic dark matter particles $\bar{X}$ is given by
\begin{equation}
\Omega_X h^2 = {1.07 \times 10^9 \over \text{GeV}} \left({(n+1) x_f^{n+1} \over \sqrt{g_*} \sigma_0 M_{Pl}}\right),
\end{equation}
where we have used the fact that $g_{*,S}(T) = g_*(T)$ in our case (all particle species have a common temperature).  
We will use the present day dark matter energy density to be $\Omega_{DM} h^2 = 0.112 \pm 0.006$~\cite{Cosmo} 
for our numerical study and in our case $n=0$.

It is important to mention that in this model we have the following free parameters:
\begin{eqnarray}
M_{Z_{BL}}, \ g_{BL}, \  M_{\bar{X}}, \ n_{BL},  
\end{eqnarray}
together with MSSM parameters $M_{\tilde{f}_i}$, and $\tan \beta$. Our results are not very sensitive 
to the values of $\tan \beta$ since the annihilation cross section in our study is basically 
independent of this parameter. In order to illustrate our main 
idea we will show the numerical results in simplified models where the dark matter candidate can 
annihilate into two MSSM sleptons when these channels are available. When the dark matter 
mass is below the slepton mass one can have the annihilation into two SM fermions at the one-loop level.
In this Letter we will focus on the simplest possibility which corresponds to the case when $\bar{X}$ is 
always heavier than the sleptons in the MSSM and the squarks are much heavier.  
Before we do the numerical analysis it is important to understand the spectrum of sfermions in this theory. 
This aspect of the theory has been studied in Ref.~\cite{FileviezPerez:2012mj}. Here we will assume for simplicity 
that the lightest sfermions are the sleptons, and their masses are given by
\begin{widetext}
\begin{eqnarray}
M_{\tilde{\nu}_i}^2 &=& M_{\tilde{L}_i}^2 \ + \ \frac{1}{2} M_{Z}^2 \cos 2 \beta - \frac{1}{2} M_{Z_{BL}}^2, \\
M_{\tilde{e}_i}^2 &=& M_{\tilde{L}_i}^2 \ + \  M_{e_i}^2 \ - \ \left(  \frac{1}{2} - \sin \theta_W^2 \right) M_{Z}^2 \cos 2 \beta - \frac{1}{2} M_{Z_{BL}}^2. 
\end{eqnarray} 
\end{widetext}
Notice that the rest of the fields with positive B-L are heavier due to the positive contribution from the B-L D-term.
\begin{figure}[h]
\scalebox{0.85}{\includegraphics{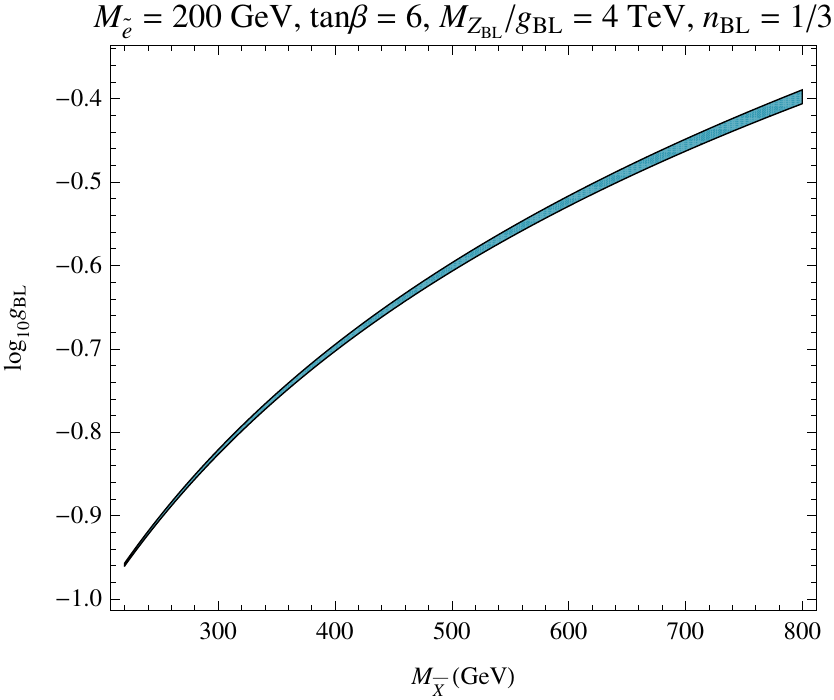}}
\caption{Allowed values for the gauge coupling $g_{BL}$ and $M_{\bar{X}}$ when $\tan \beta=6$, $M_{Z_{BL}}/g_{BL}=4$ TeV, 
$M_{\tilde{e}}=200$ GeV and $n_{BL}=1/3$. Here we assume the annihilation to only one family of sleptons. 
The black lines produce $\Omega h^2 = 0.112$ while the blue region represents $\Omega h^2 \leq 0.112$.}
\end{figure}
\begin{figure}[h]
\scalebox{1.0}{\includegraphics{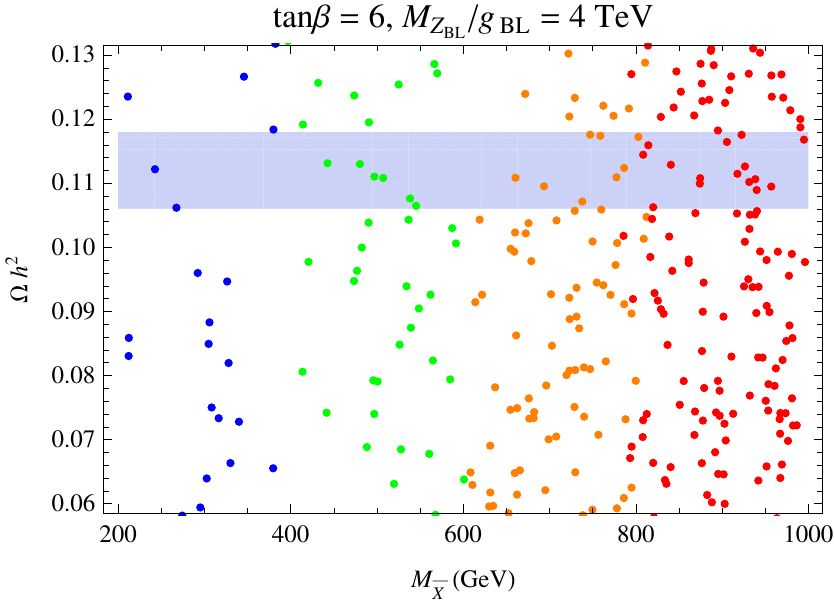}}
\caption{Values for the relic density vs the dark matter mass $M_{\bar{X}}$, when $\tan \beta=6$, $n_{BL}=1/3$, and $M_{Z_{BL}}/g_{BL}=4$ TeV. 
Here the blue dots correspond to the solutions for $g_{BL}=0.1 \div 0.2$, the green dots are for $g_{BL}=0.2 \div 0.3$, the orange dots are for $g_{BL}=0.3 \div 0.4$, and the red dots are for $g_{BL}=0.4 \div 0.5$. 
The slepton mass changes between 100 GeV and the dark matter mass.}
\end{figure}
In Fig.~1 we show the allowed values for the gauge coupling $g_{BL}$ and DM mass $M_{\bar{X}}$ 
when $\tan \beta=6$, $M_{Z_{BL}}/g_{BL}=4$ TeV, $M_{\tilde{e}}=200$ GeV and $n_{BL}=1/3$. 
Here we assume a simplified model where the annihilation is only possible to one family of sleptons. 
Notice that for this type of scenario the gauge coupling has to be change between $10^{-1}$ and $10^{-0.4}$, 
in order to achieve the relic density consistent with cosmological observations. As we will discuss later, 
this type of scenario is allowed by the constraints coming from direct detection experiments, 
which we discuss in detail in the next section.

In order to have a better idea of how to achieve the right relic density we show in Fig.~2 the values for the relic density when changing the dark matter 
mass $M_{\bar{X}}$, when $\tan \beta=6$, $n_{BL}=1/3$, and $M_{Z_{BL}}/g_{BL}=4$ TeV. Here the blue dots correspond to the solutions for $g_{BL}=0.1 \div 0.2$, the green dots are for $g_{BL}=0.2 \div 0.3$, the orange dots are for $g_{BL}=0.3 \div 0.4$, and the red dots are for $g_{BL}=0.4 \div 0.5$. We also scan over the slepton mass between 100 GeV and the dark matter mass. Notice that we find many solutions 
which are consistent with relic density bounds when the gauge coupling is between $0.3$ and $0.5$. 

It is easy to understand the results presented in Fig.~1 and Fig.~2. When the gauge coupling is small 
or we increase the dark matter mass we suppress the annihilation cross section, we can achieve 
the relic density allowed by the experiments. The only way to achieve solutions when the gauge coupling 
is close to one is to suppress the phase space choosing a small splitting between the slepton mass and 
the dark matter mass. Notice that the annihilation through the $Z_{BL}$ is suppressed in these scenarios 
because the B-L gauge boson is very heavy and the cross section is p-wave suppressed. Also we can 
have other annihilation channels into two quarks at one-loop level but these are also suppressed by the squark masses.  
%
\section{IV. Predictions for DM Direct Detection}
%
The couplings of the $Z_{BL}$ to quarks and the dark matter candidate, $\bar{X}$, can lead to a 
potentially sizable spin-independent elastic scattering cross section between dark matter and nuclei. 
The cross section in this case is given by
\begin{eqnarray}
\sigma^{\rm SI}  = \frac{M^2_{\bar{X}} m^2_N}{\pi (M_{\bar{X}} + m_N)^2} \bigg[Z \ f_p \ + \  (A-Z) \  f_n \bigg]^2,  
\end{eqnarray}
where $A$ and $Z$ are the atomic mass and atomic number of the target nucleus and $f_{(p,n)}$ 
are the effective couplings to protons and neutrons:
\begin{eqnarray}
f_p &=& \frac{g_{\bar{X} \bar{X} Z_{BL}} (2 \ g_{u u Z_{BL}} \ + \  g_{dd Z_{BL}})}{M^2_{Z_{BL}}},\\
f_n &= & \frac{g_{\bar{X} \bar{X} Z_{BL}} (g_{uu Z_{BL}} \ + \ 2 \ g_{dd Z_{BL}})}{M^2_{Z_{BL}}}. 
\end{eqnarray}
Here, we have used $g_{\bar{X} \bar{X} Z_{BL}}$ and $g_{qqZ_{BL}}$ to denote the effective 
coupling strengths of the respective vertices. For any quark we have $g_{qqZ_{BL}}=g_{BL}/6$ and 
$g_{\bar{X} \bar{X} Z_{BL}}=n_{BL} \, g_{BL}/2$. Now, using the relations above we can write the 
dark matter nucleon cross section as
\begin{widetext}
\begin{equation}
\sigma^{\rm SI}_{\bar{X}n }  (\rm{cm}^2)= (1.2 \times 10^{-40}\ {\rm cm^2}) \times g_{BL}^4 n_{BL}^2  \times \left(\frac{500 \ {\rm GeV}}{M_{Z_{BL}}}\right)^4 \times \left(\frac{\mu}{1 \ {\rm GeV}}\right)^2,
\end{equation}
\end{widetext}
where $\mu = M_{\bar{X}} m_n /(M_{\bar{X}} + m_n)$, and $m_n$ is the nucleon mass.
It is well-known that the dark matter spin-independent elastic cross sections are tightly constrained by the Xenon100 experimental results~\cite{Aprile}.
In Fig.~3 we show the numerical values for the elastic DM-nucleon cross section as a function of the dark matter 
mass $M_{\bar{X}}$. Here we use different values for the ratio  $M_{Z_{BL}}/(g_{BL}\sqrt{n_{BL}})$ 
and show the bounds from Xenon10~\cite{Angle:2011th} and Xenon100~\cite{Aprile} experiments.
\begin{figure}[h]
\scalebox{1.0}{\includegraphics{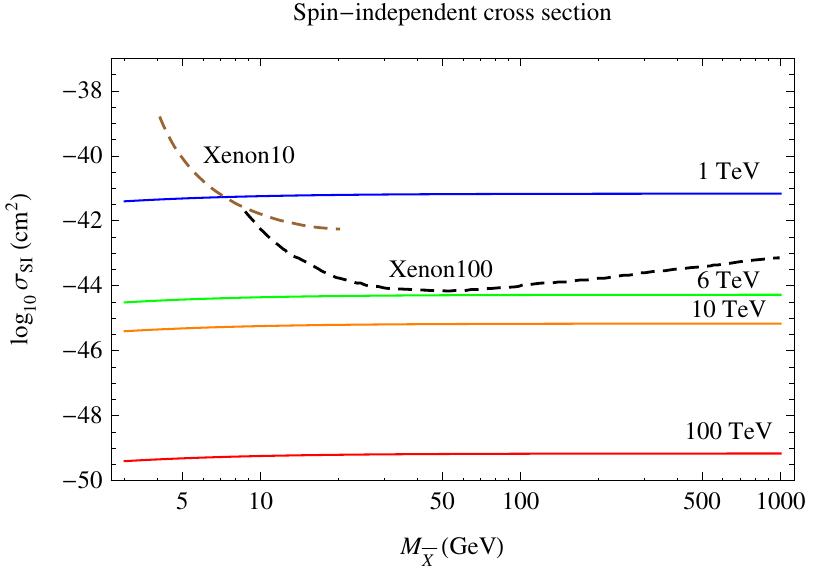}}
\caption{Values for the spin-independent elastic DM-nucleon cross section for  a few different ratios 
$M_{Z_{BL}}/(g_{BL}\sqrt{n_{BL}})$: $1 \ \rm TeV$ (blue), $6 \ \rm TeV$ (green), $10 \ \rm TeV$ (orange), and $100 \ \rm TeV$ (red). 
The black dashed line is the exclusion limit from Xenon100 and the brown dashed line is the exclusion limit from Xenon10. Note that the 6 TeV line in this plot is consistent with our earlier choice of $M_{Z_{BL}}/g_{BL}= 4$ TeV when $n_{BL}=1/3$.}
\end{figure}
The best limits on our model come from Xenon100, which for $M_{\bar{X}} \gtrsim 30 \ \rm GeV$ rules out most of the region 
$M_{Z_{BL}}/(g_{BL}\sqrt{n_{BL}}) < 6 \ \rm TeV$. On the other hand, ratios $M_{Z_{BL}}/(g_{BL}\sqrt{n_{BL}})$ 
as low as $1 \ \rm TeV$ are allowed for light dark matter masses, $M_{\bar{X}} \lesssim 8 \ \rm GeV$.
It is important to mention that the collider bound on the B-L gauge boson is about $M_{Z_{BL}}/g_{BL} > 3$ TeV.
Then, we can say that the dark matter experiment Xenon100 sets a strong bound on the gauge boson mass 
if $n_{BL}$ is not very small.

In Fig.~4 we show the correlation between the values for the spin-independent cross section and the dark 
matter relic density when $M_{\tilde{e}}=100$ GeV, $\tan \beta=6$, $n_{BL}=1/3$, 
$0.1 \ \rm{TeV} \leq  M_{Z_{BL}} \leq  10 \ \rm{TeV}$, and $0.1 \leq g_{BL} \leq 1$, 
for different values of the dark matter mass. Then, in this way we can see that there are not many allowed 
solutions by the relic density constraints assuming the mentioned values of the free parameters.  
Since the range of the parameter space is quite representative we can say that it is not easy to find solutions 
in agreement with the experiments. 
\begin{figure}[h]
\scalebox{1.0}{\includegraphics{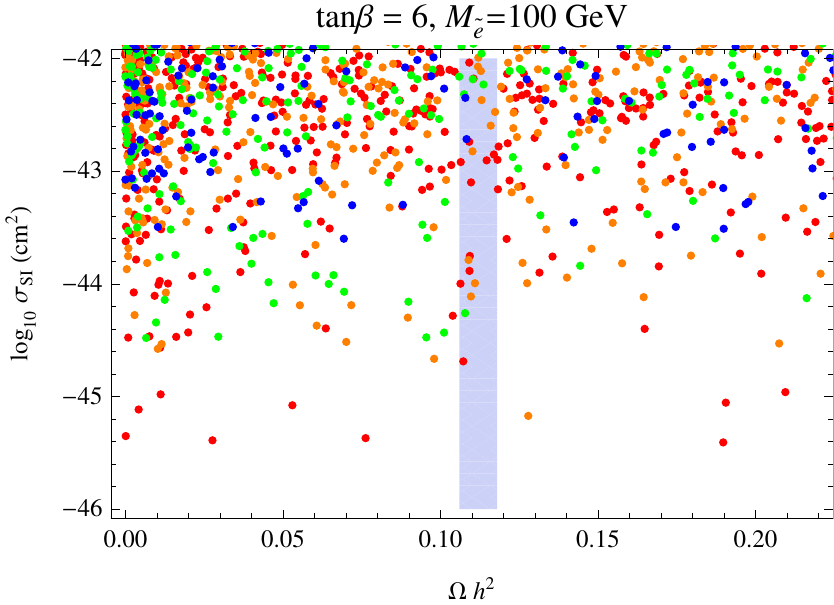}}
\caption{Values for the cross section allowed by the relic density constraints when the slepton mass is 100 GeV, $\tan \beta=6$, $n_{BL}=1/3$, 
$0.1 \ \rm{TeV} \leq  M_{Z_{BL}} \leq  10 \ \rm{TeV}$, and $0.1 \leq g_{BL} \leq 1$. Blue, green, orange, and red dots correspond to $M_{\bar{X}} = 120, 200, 300$, and $400$ GeV, respectively.}
\end{figure}
If we think about the testability of this model for dark matter one can imagine a very optimistic scenario where we can know the parameters 
$M_{Z_{BL}}$, $g_{BL}$, $M_{\tilde{e}}$, and $\tan \beta$ from the Large Hadron Collider or future collider experiments. 
Then, we could get the rest of the parameters $n_{BL}$ and the dark matter mass $M_{\bar{X}}$ using the constraints from relic density 
and direct detection experiments. For example, suppose that in a dark matter experiment such as Xenon100 you find 
a signal which corresponds to a cross section of $10^{-45} \  \rm{cm}^2$. If the collider experiments measure say, $g_{BL}=0.3, \tan\beta=6, M_{Z_{BL}}=2$ TeV and $M_{\tilde{e}}=200$ GeV and we require $\Omega h^2 \leq 0.112$, this corresponds to $n_{BL}=0.54$ and $986$ GeV $\leq M_{\bar{X}} \leq 1014$ GeV.  The bounds of this inequality achieve 
$\Omega h^2 = 0.112$. In this way we could think about the testability of this scenario. Of course, that the discovery of supersymmetry and of a B-L gauge boson is 
crucial to start thinking about it.  
%
\section{Summary and Outlook}
In this Letter we have investigated a simple scenario for the cold dark matter 
in the universe where the sector responsible for dark matter has ``exact" supersymmetry 
before symmetry breaking. In order to achieve this type of scenario we assume 
that supersymmetry breaking is mediated as in ``gauge mediation", where the messenger 
fields only have quantum numbers of the visible sector, and the soft terms induced by gravity 
are very small. The SM singlet fields in the dark sector do not get large soft 
terms from gravity mediation and we can say that supersymmetry is a 
good symmetry in the dark sector.

In order to illustrate our idea we consider the case where in the visible sector 
we have the simplest B-L extension of the minimal supersymmetric standard model 
while the dark sector is composed of two chiral superfields with B-L quantum numbers. 
We have found that in this case the dark matter candidate is the lightest scalar field 
in the dark sector and the B-L D-term induces a mass splitting after the symmetry is broken. 

We noticed that the dark matter candidate is stable even if R-parity is spontaneously broken 
in the visible sector. Since the link between the visible and dark sectors is through the B-L 
gauge force, the dark matter annihilates mainly into two sfermions when these channels are 
available. We have shown the allowed parameter space by the relic density and direct detection 
experiments in simplified scenarios where the annihilation is mainly into two sleptons.
In the case when the dark matter candidate is below 100 GeV, the DM annihilation is mainly into 
two fermions at the one-loop level where inside the loops you have the sfermions and gauginos. 
The details of the scenario for light dark matter and the annihilation into photons will be 
investigated in a future publication~\cite{Long-paper}.

\vspace{1.0cm}
{\textit{Acknowledgments}}:
{\small We would like to thank Manuel Drees, Pran Nath, Stefano Profumo, Sogee Spinner and Mark B. Wise for comments and discussions.
The work of P.F.P. has been supported by the James Arthur Fellowship, CCPP, New York University. 
The work of JMA and BF was supported in part by the U.S. Department of Energy under contract No.
DE-FG02-92ER40701.}


\begin{thebibliography}{000}

\bibitem{Drees:2012sz}
  M.~Drees and G.~Gerbier,
  ``Mini-Review of Dark Matter: 2012,'' in 
   [Particle Data Group Collaboration],
  Phys.\ Rev.\ D {\bf 86} (2012) 010001.

  
\bibitem{Goldberg:1983nd}
  H.~Goldberg,
  ``Constraint on the Photino Mass from Cosmology,''
  Phys.\ Rev.\ Lett.\  {\bf 50} (1983) 1419
   [Erratum-ibid.\  {\bf 103} (2009) 099905].
  
\bibitem{Ellis:1983ew}
  J.~R.~Ellis, J.~S.~Hagelin, D.~V.~Nanopoulos, K.~A.~Olive and M.~Srednicki,
  ``Supersymmetric Relics from the Big Bang,''
  Nucl.\ Phys.\ B {\bf 238} (1984) 453.
  

\bibitem{Jungman:1995df}
  G.~Jungman, M.~Kamionkowski and K.~Griest,
  ``Supersymmetric dark matter,''
  Phys.\ Rept.\  {\bf 267} (1996) 195
  [hep-ph/9506380].
  
\bibitem{Buchmuller:2009fm}
  W.~Buchmuller,
  ``Gravitino Dark Matter,''
  AIP Conf.\ Proc.\  {\bf 1200} (2010) 155
  [arXiv:0910.1870 [hep-ph]].
   
\bibitem{Krauss:1988zc}
  L.~M.~Krauss and F.~Wilczek,
  ``Discrete Gauge Symmetry in Continuum Theories,''
  Phys.\ Rev.\ Lett.\  {\bf 62} (1989) 1221.
  
\bibitem{Martin:1992mq}
  S.~P.~Martin,
  ``Some simple criteria for gauged R-parity,''
  Phys.\ Rev.\ D {\bf 46} (1992) 2769
  [hep-ph/9207218].
  
\bibitem{Aulakh:1999cd}
  C.~S.~Aulakh, A.~Melfo, A.~Rasin and G.~Senjanovic,
  ``Seesaw and supersymmetry or exact R-parity,''
  Phys.\ Lett.\ B {\bf 459} (1999) 557
  [hep-ph/9902409].
  
\bibitem{Aulakh:2000sn}
  C.~S.~Aulakh, B.~Bajc, A.~Melfo, A.~Rasin and G.~Senjanovic,
  ``SO(10) theory of R-parity and neutrino mass,''
  Nucl.\ Phys.\ B {\bf 597} (2001) 89
  [hep-ph/0004031].
  
\bibitem{Babu:2008ep}
  K.~S.~Babu and R.~N.~Mohapatra,
  ``Minimal Supersymmetric Left-Right Model,''
  Phys.\ Lett.\ B {\bf 668} (2008) 404
  [arXiv:0807.0481 [hep-ph]].
  
\bibitem{Feldman:2011ms}
  D.~Feldman, P.~Fileviez Perez and P.~Nath,
  ``R-parity Conservation via the Stueckelberg Mechanism: LHC and Dark Matter Signals,''
  JHEP {\bf 1201} (2012) 038
  [arXiv:1109.2901 [hep-ph]].
  
\bibitem{Perez:2011zx}
  P.~Fileviez Perez, S.~Spinner and M.~K.~Trenkel,
  ``Testing the Mechanism for the LSP Stability at the LHC,''
  Phys.\ Lett.\ B {\bf 702} (2011) 260
  [arXiv:1103.3824 [hep-ph]].
  
\bibitem{Basso:2012gz}
  L.~Basso, B.~O'Leary, W.~Porod and F.~Staub,
  ``Dark matter scenarios in the minimal SUSY B-L model,''
  arXiv:1207.0507 [hep-ph].

\bibitem{O'Leary:2011yq}
  B.~O'Leary, W.~Porod and F.~Staub,
  ``Mass spectrum of the minimal SUSY B-L model,''
  JHEP {\bf 1205} (2012) 042
  [arXiv:1112.4600 [hep-ph]].
  
\bibitem{FileviezPerez:2011kd}
  P.~Fileviez Perez, S.~Spinner and M.~K.~Trenkel,
  ``The LSP Stability and New Higgs Signals at the LHC,''
  Phys.\ Rev.\ D {\bf 84} (2011) 095028
  [arXiv:1103.5504 [hep-ph]].
    
\bibitem{Barger:2008wn}
  V.~Barger, P.~Fileviez Perez and S.~Spinner,
  ``Minimal gauged U(1)(B-L) model with spontaneous R-parity violation,''
  Phys.\ Rev.\ Lett.\  {\bf 102} (2009) 181802
  [arXiv:0812.3661 [hep-ph]].
  
  
\bibitem{FileviezPerez:2012mj}
  P.~Fileviez Perez and S.~Spinner,
  ``The Minimal Theory for R-parity Violation at the LHC,''
  JHEP {\bf 1204} (2012) 118
  [arXiv:1201.5923 [hep-ph]].
  
\bibitem{Cosmo}
  O.~Lahav and A. R. Liddle,
  ``The Cosmological Parameters,'' \\
  http://pdg.lbl.gov/2012/reviews/rpp2012-rev-cosmological-parameters.pdf

    
\bibitem{Aprile}
  E.~Aprile {\it et al.}  [XENON100 Collaboration],
  ``Dark Matter Results from 100 Live Days of XENON100 Data,''
  Phys.\ Rev.\ Lett.\  {\bf 107}, 131302 (2011)
  [arXiv:1104.2549 [astro-ph.CO]].
 
\bibitem{Angle:2011th}
  J.~Angle {\it et al.}  [XENON10 Collaboration],
  ``A search for light dark matter in XENON10 data,''
  Phys.\ Rev.\ Lett.\  {\bf 107} (2011) 051301
  [arXiv:1104.3088 [astro-ph.CO]].
  
  

\bibitem{Long-paper}
  J. M. Arnold, P.~Fileviez Perez and B. Fornal, in preparation.

\end{thebibliography}
\end{document}